# Mechanistic Transition from Screw to Edge Dislocation Glide Enhances High-Temperature Strength in Refractory Complex Concentrated Alloys

Tamanna Zakia[1], Ayeman Nahin[1], Dunji Yu[2], Jacob Pustelnik[1], Juntan Li[3], Mason Kincheloe[1], Lia Amalia[2,3], Yan Chen[2], Peter K. Liaw[3], Haixuan Xu[3], Mingwei Zhang[1, *]

[1]Department of Materials Science and Engineering, University of California, One Shields Ave., Davis, CA, 95616, USA.
[2] Neutron Scattering Division, Oak Ridge National Laboratory, Oak Ridge, TN, 37830
[3] Department of Materials Science and Engineering, University of Tennessee, Knoxville, TN, 37996, USA

*Corresponding Author: Mingwei Zhang (mwwzhang@ucdavis.edu)

## Abstract

The strength of body-centered cubic materials is traditionally known to be governed by screw dislocations. However, recent findings reveal that in certain refractory complex concentrated alloys, edge dislocations can instead control strength. This work integrates high-temperature mechanical testing, in-situ neutron scattering during heating and tension, scanning transmission electron microscopy, and molecular dynamics simulations to uncover the mechanism behind this behavior. In the Nb-Ta-Ti-V system, increasing the V content, due to its smaller atomic size, induces substantial atomic misfit that raises the glide barrier for edge dislocations relative to screw dislocations. This effect drives a gradual transition from screw to edge dislocation–controlled deformation, leading to markedly enhanced strength at elevated temperatures.

# Introduction

High-temperature structural applications, such as aerospace-propulsion systems and hot-gas-path components in turbine systems for electricity generation, demand materials capable of sustaining extreme conditions involving elevated temperatures and high mechanical stresses (*1*). Although conventional Ni-based superalloys have provided reliable service for decades, their application is fundamentally constrained to temperatures up to ~1473 K as they approach their melting points (*2, 3*). This limitation has motivated exploration of alternative alloys with higher melting points and improved high-temperature mechanical properties. Refractory metals, including Ti, V, Cr, Zr, Nb, Mo, Hf, Ta, W, and their alloys have emerged as natural candidates for next-generation high temperature materials due to their exceptionally high melting temperatures. In recent years, alloy design strategies have advanced from simple binary or ternary systems to refractory complex concentrated alloys (RCCAs), which are sometimes also known as refractory high entropy alloys (RHEAs) (*2–5*). RCCAs are composed of three or more principal refractory elements in near-equiatomic or in high concentrations. They often form single-phase BCC solid solutions but sometimes also give rise to secondary phases such as hexagonal-close-packed (HCP), ordered B2, or Laves phases (*6, 7*). Multiphase RCCAs typically exhibit higher yield strength at room temperature than their single-phase counterparts, but their ability to retain strengths at elevated temperatures is generally inferior to that of single-phase RCCAs as the secondary phases can dissolve at higher temperatures (*8*).

To examine the applicability of RCCAs at elevated temperatures, it is critical to understand how their yield strengths vary with temperature. Extensive experimental evidence has proved that the temperature dependence of yield strength in refractory metals and alloys generally encompasses three regimes: a trend of decreasing strength with increasing temperature at both low and high temperature extremes while exhibiting a temperature-independent plateau at intermediate temperatures (*8–12*). Understanding the scientific origin of this plateau behavior is crucial, as it provides the foundation of designing new alloys with improved strength retention to still higher temperatures. In refractory BCC metals and conventional alloys, the plateau typically extends from ~0.1–0.2 $T_m$ to ~0.5–0.6 $T_m$, where $T_m$ is the absolute melting temperature. This plateau was rationalized by the athermal contributions of flow stress in BCC metals (*9–16*). In pure BCC metals, strength and plastic deformation are primarily governed by the glide of screw dislocations, which must overcome both short-range and long-range barriers. The short-range barriers, associated with the nucleation of mobile kink pairs on relatively immobile screw dislocations, are highly temperature-sensitive and can be overcome by thermal activation of ~0.1–0.2 $T_m$. Once these barriers are overcome, strength becomes controlled by the less temperature-sensitive long-range barriers, including dislocation-dislocation interactions and cross-slip processes, which enable strength retention up to a critical temperature. Beyond ~0.5–0.6 $T_m$, these long-range barriers are overcome by diffusion-controlled mechanisms and dynamic recrystallization, leading to a rapid decline in strength with increasing temperature (*17, 18*).

In RCCAs, the intermediate plateau typically spans ~0.2–0.4 $T_m$, which is narrower and steeper than that observed in pure BCC metals (*9*). The steeper slope indicates a stronger temperature sensitivity compared to conventional BCC metals. Moreover, the fact that the plateau begins at temperatures higher than those associated with kink-pair nucleation and ends below the onset of diffusion-mediated processes and dynamic recrystallization suggests that the origin and underlying mechanism of the plateau in RCCAs are fundamentally different from those in pure BCC metals and dilute alloys. To address the effect of concentrated solid solutions, screw dislocation strengthening theories (*19*) have been developed for RCCAs that suggests screw dislocations in concentrated solid solutions naturally adopt a kinked structure, unlike pure BCC metals, and strength is controlled by the formation and dragging of dislocation dipoles at low temperatures and jogs at high temperatures (*20*). However, these efforts could not explain the yield strength plateau at intermediate temperatures because the considered controlling mechanisms based on screw dislocation activities are strongly thermally activated and a monotonous decrease in strength as a function of temperature is predicted.

Recently, it has been suggested that in some RCCAs, strength can be controlled by edge dislocations due to a large energy barrier against their motion within the unique potential energy landscapes of these RCCAs (*21*). Specifically, in CrMoNbV and NbTaTiV, the high strength and effective strength retention at elevated temperatures are attributed to edge dislocations, where {110} <111> edge dislocations were frequently observed using a combination of in-situ neutron scattering and transmission electron microscopy characterization. Of particular interest is the NbTaTiV RCCA, which exhibits exceptional mechanical properties, showing over 40% tensile elongation at room temperature (*22*), a yield strength of ~1 GPa, and maintaining more than 600 MPa yield strength at 1173 K (*21*). This combination represents an excellent strength–ductility synergy among single-phase BCC RCCAs and highlights the need to fundamentally understand the potential transition from screw dislocation glide-controlled strength, which governs strength in conventional refractory alloys and most RCCAs, to edge dislocation glide-controlled strength, which appears to dominate in NbTaTiV and CrMoNbV. It is previously established that in the ductile NbTaTiHfZr system (also known as the Senkov alloy (*23*, *24*)) , plastic deformation is mainly controlled by the glide of screw dislocations(*24–27*) . In the $Nb_{45}Ta_{25}Ti_{15}Hf_{15}$ RCCA, glide and intersection of screw dislocation dominates the onset of plastic deformation, whereas coordinated slip of edge dislocations form kink bands during a later stage of plastic deformation that provide exceptional fracture toughness over 253 $MPa.m^{1/2}$ . At 1173K, tensile creep deformation in the $Nb_{45}Ta_{25}Ti_{15}Hf_{15}$ RCCA was reported to be controlled by cross-kink collisions and jog dragging from screw dislocations, where kink band formation appeared to be missing at lower strain rates (*28*). As a result, the creep rates of $Nb_{45}Ta_{25}Ti_{15}Hf_{15}$ can be completely rationalized by the screw dislocation-mediated Rao-Suzuki model (*29*). This list can be further extended to the ternary alloy, NbTiZr, which is known to be controlled by glide of screw dislocations and kink formation on varying {110} planes (*30*, *31*). At 77 K, deformation occurred by screw dislocation slip only, with long, straight screw dislocations being detected. As the temperature increases, cross-kinks/jogs as well as the formation of dislocation debris (in the form of vacancy and interstitial loops) have been identified. Most interestingly, plastic deformation in NbTaTi is also shown to be screw dominated, where the mobility of edge dislocations is predicted to be much higher than that of screw dislocations even up to elevated temperatures (*10*, *20*, *32*).

This contrast raises a critical question regarding why V-containing NbTaTiV exhibits edge dislocation glide–controlled strength while simultaneously achieving exceptional tensile ductility, high elevated-temperature strength, and superior strength retention at elevated temperatures, whereas the strengths of non-V-containing NbTaTi-based RCCAs are controlled by screw dislocations. The mechanistic origin that facilitates a transition from screw to edge dislocation-controlled strength can be hypothesized through the alloying of V, which possesses a significantly smaller atomic volume compared to the other elements. The large misfit volume brought by a significant V concentration can result in severe lattice distortion (*33*), where in prior studies, the high strength of NbTaTiV has been attributed to the lattice distortion that created significant resistance to the motion of edge dislocation (*13–16*). Recent atomistic simulations of BCC RCCAs report the temperature-dependent crossover in edge and screw dislocation flow stresses and show that edge dislocations can control the strength in MoNbTi and NbMoTaW at elevated temperatures (*34*). Notably, these simulated alloys also possess noticeable atomic mismatch, similar to the NbTaTiV system. However, there is a lack of comprehensive experimental evidence and mechanistic insights to date that can establish the relationship between lattice distortion and the shift from screw to edge dislocation glide-control of strength. In addition, the impact of this mechanistic transition on the temperature dependence of yield strength, and particularly on the pronounced yield-strength plateau observed in the NbTaTiV system (*21*), remains unclear.

The motivation of the present work is therefore to elucidate the role of V in promoting the transition from screw to edge dislocation-controlled strength at elevated temperatures by introducing a gradient of V concentration in the pseudobinary $(NbTaTi)_{1-x}V_x$ system. It is hypothesized that this transition can be rationalized by the atomic size misfit caused by V, which generates hydrostatic stress fields that interact primarily with edge dislocations and elevate the edge dislocation–controlled strength above that governed by screw dislocations. The mechanistic transition is shown by the combination of meticulously controlled alloy processing to produce samples with similar grain sizes, interstitial contents, and a homogeneous BCC single-phase structure, elevated-temperature tensile testing under high vacuum, dislocation analyses through in-situ heating/tension neutron scattering experiments, scanning transmission electron microscopy, as well as molecular dynamics (MD) simulations that compute the flow stresses for screw and edge dislocations at the onset of plasticity.

**Results**
**Initial microstructure and elevated-temperature mechanical properties of the $(NbTaTi)_{1-x}V_x$ RCCAs**

The initial microstructure of the four alloys investigated in this study, NbTaTi, $(NbTaTi)_{92}V_8$, $(NbTaTi)_{84}V_{16}$, and NbTaTiV is provided in **Fig. 1**. For simplicity, the four alloys are sometimes referred to as V0, V8, V16, and V25 according to the V content. The homogenized alloys exhibited a body-centered cubic (BCC) single phase with clean grain boundaries with no evidence of secondary phases formation in our scanning electron microscopy (SEM), energy dispersive X-ray spectroscopy (EDS), and neutron scattering analyses. The average grain sizes, EDS-measured chemical compositions, average lattice spacing, as well as the oxygen and nitrogen (O/N) interstitial impurity concentrations are provided in **Table 1.** It is observed that the lattice spacing in the $(NbTaTi)_{1-x}V_x$ system decreases with V content, which is caused by the smaller atomic size of V compared to the other elements. The annealing parameters were adjusted to ensure comparable grain sizes across all samples. In addition, all materials fabrication and processing steps were meticulously controlled in high vacuums to minimize the uptake of interstitial impurities. Therefore, any observed differences in mechanical properties across the four compositions can be attributed to the effect of V alloying, rather than to variations in Hall–Petch strengthening or solid-solution strengthening from O/N atoms.

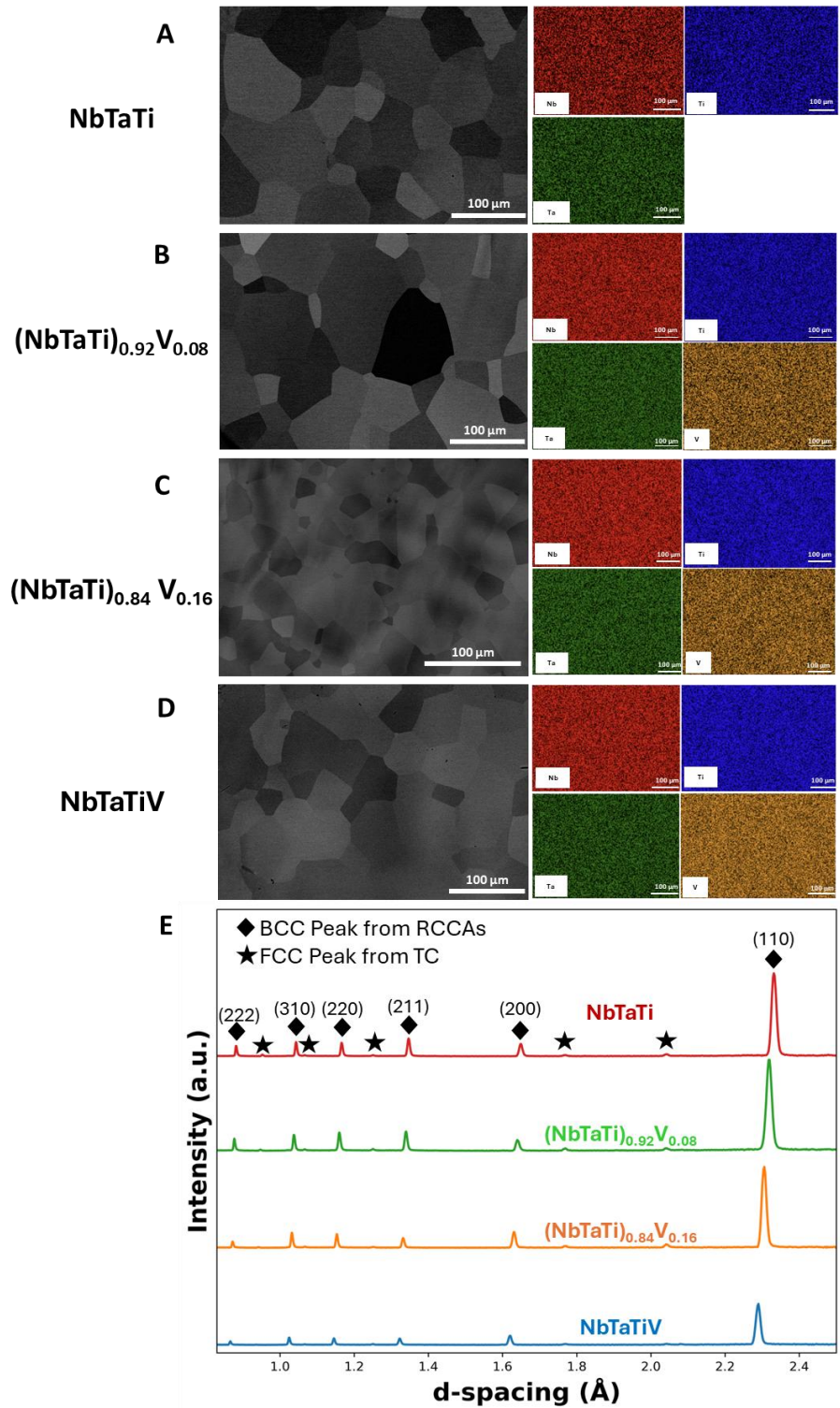

**Fig. 1 Microstructure of all the four alloys with increasing vanadium concentration.** (A-D) SEM-BSE image with corresponding EDS elemental maps of the homogenization and recrystallized alloys. (E) Neutron diffraction patterns of the stable BCC phase for the alloys (intensity in arbitrary units). Minor FCC peaks arise from the K-type thermocouple (TC) used during the measurements.

**Table 1. List of nominal compositions, grain sizes, actual average composition measured by EDS, neutron measured lattice constants, and inert gas fusion O and N interstitial ppms (in atomic ppm).**

| Nominal Composition | Grain Size (μm) | Experimental Lattice Constants (Å) | EDS Composition (Atomic %) | | | | Inert Gas Fusion Composition (appm) | |
|---|---|---|---|---|---|---|---|---|
| | | | Nb | Ta | Ti | V | O | N |
| NbTaTi | 43.21 ± 8.52 | 3.2965 | 34.3 | 33.4 | 32.3 | - | 5736.39 | 459.08 |
| $(NbTaTi)_{0.92}V_{0.08}$ | 50.39 ± 14.34 | 3.2781 | 28.91 | 29.42 | 33.74 | 7.93 | 4031.73 | 1464.76 |
| $(NbTaTi)_{0.84}V_{0.16}$ | 57.70 ± 9.40 | 3.2573 | 26.69 | 26.31 | 30.48 | 16.52 | 5865.38 | 910.8 |
| NbTaTiV | 61.98 ± 9.04 | 3.2393 | 23.74 | 22.79 | 27.86 | 25.61 | 5333.4 | 531.76 |

**Fig. 2A** shows the temperature-dependent 0.2% offset yield strength for all four compositions. The tensile yield strengths for NbTaTi and NbTaTiV obtained in this study compare favorably with the compressive yield strengths previously obtained by Coury et al. (*10*) and Lee et al. (*21*) respectively. It is immediately evident that the V addition systematically enhances the yield strength of the NbTaTi system across the entire temperature range. To relate the variations of the yield strength across the four alloys to lattice distortion, intrinsic lattice distortion ($\overline{u^D}$) as a function of the atomic fraction of V is presented in **Fig. S1**. The continuous curve represents the theoretical prediction, calculated following the quantitative lattice distortion model developed by Lee *et al.* (*35*). Detailed methodology for this calculation can be seen in **Supplementary Text**. It is evident that in the pseudo-binary $(NbTaTi)_{1-x}V_x$ system, a consistent trend of increasing lattice distortion with V addition can be obtained until approximately 50 at. % V. Lattice distortion then gradually tapers down to zero when pure V is reached. In addition, the rate of increase in lattice distortion decreases with V concentration below 50 at. %, where the choice of V0, V8, V16, V25 is sensible in that it can capture the most significant differences in lattice distortion caused by V alloying.

The significant strengthening behavior caused by V alloying correlates well with the increased lattice distortion and the development of local atomic-level strain fields due to the atomic misfit volume induced by V (*33*). Such distortion effectively impedes dislocation glide and leads to the pronounced solid-solution strengthening, especially in the intermediate temperature regime. **Fig. 2B** further provides the temperature-dependent elastic modulus normalized yield strength and melting point (in this case, the liquidus temperature $T_{liquid}$ is used) normalized temperature. It is observed that NbTaTi exhibits weak plateau behavior with increasing temperature, while V-containing alloys show distinct thermal and athermal regimes. At lower temperatures (<700 K, or near 0.2 $T_{liquid}$), all alloys experience a sharp decrease in strength which corresponds to the onset of thermally activated dislocation motion. Between approximately 600 and 1000 K (or near 0.2-0.4 $T_{liquid}$), the V-containing alloys display a distinct plateau region, where the yield strength remains nearly constant despite the increasing temperature. Notably, the extent of this plateau increases with V content, and NbTaTiV exhibits the widest plateau region up to 1173 K (or approximately 0.45 $T_{liquid}$). It is further noted that V8, V16, and V25 consistently exhibit the yield strength anomaly (YSA) behavior, where the yield strength increases with increasing temperature near 773 K. Interestingly, the yield strengths of the four compositions beyond the intermediate-temperature plateau converge at approximately 1500

K. This convergence suggests that, at sufficiently high temperatures, dislocation glide may not be relevant in controlling the strength of RCCAs, where the effect of solid solution hardening is massively reduced. Instead, other diffusion-mediated deformation mechanisms, such as dislocation climb, dynamic recrystallization, and grain boundary sliding, can dominate the thermally activated processes and effectively overcome the barriers associated with dislocation glide and solute interactions. This leads to comparable strength levels among all compositions regardless of V content. The mechanistic study of these RCCAs in the ultrahigh temperature regime that are not potentially controlled by thermally activated dislocation glide is beyond the scope of the current investigation.

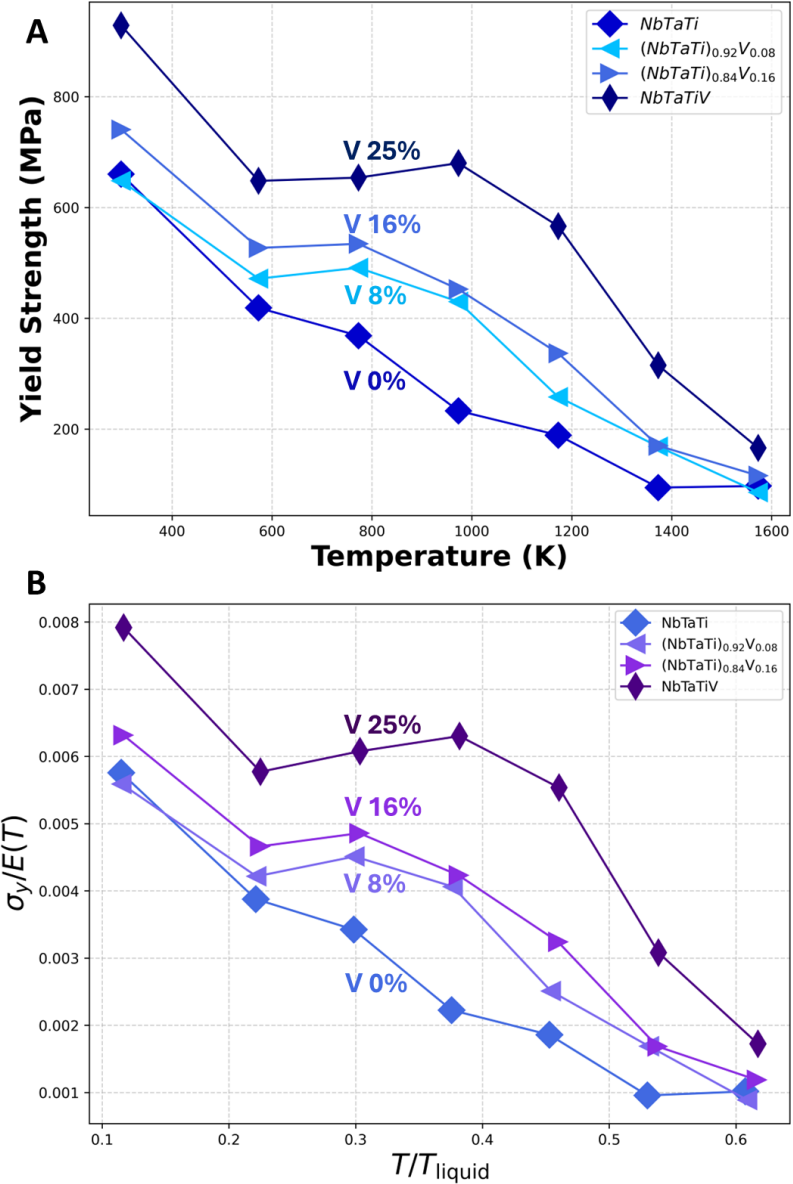

**Fig. 2 Temperature-dependent yield strength.** (A) 0.2% offset yield strength as a function of temperature for all four compositions. (B) temperature-dependent elastic modulus normalized 0.2% offset yield strength as a function of melting point normalized temperature for all four compositions.

# Transition from screw dislocation glide to edge dislocation glide-controlled strength in (NbTaTi)$_{1-x}$V$_x$ RCCAs

With the significant knowledge gained from the effect of V-alloying on the temperature dependence of yield strength, it is paramount to understand whether V can impose a transition in dislocation characters that control the yield strength of the material at high temperatures. Neutron scattering is a powerful technique to analyze representative bulk-averaged dislocation characters that control the yield strength of the RCCAs. In this study, a neutron beam as large as 8 mm x 2.5 mm was used, which can cover almost the entire gauge section of the in-situ heating/tensile specimen with full penetration of the specimen thickness. A dislocation character analysis can be performed using neutron diffraction data collected during high-temperature tensile testing near yield, based on the modified Williamson–Hall method following Lee et al. (*36*). Modified Williamson-Hall plot considers both grain size and strain effects and can be expressed as

$$\Delta K^2 = \left(\frac{0.9}{D}\right)^2 + \frac{\pi A^2 b^2}{2} \rho (K^2 C_{hkl}) + O(K^2 C_{hkl})^2, \qquad (1)$$

where $\Delta K$ represents the diffraction peak broadening, $D$ is the grain size, $A$ is a constant denoting the effective outer cutoff radius of dislocations, $b$ is the magnitude of the Burgers vector, $\rho$ is the dislocation density, $K = \frac{1}{d}$ is the inverse d-spacing, $O$ is a noninterpreted higher order term, and $C_{hkl}$ is the dislocation contrast factor. $C_{hkl}$ is obtained using the ANIZC software (*37*) and can be defined as

$$C_{hkl} = C_{h00}\left[1 - q\left(\frac{h^2k^2+k^2l^2+h^2l^2}{(h^2+k^2+l^2)^2}\right)\right]. \qquad (2)$$

Considering the single crystal elastic constant, the values of $q$ and $C_{h00}$ have been determined as 2.2244 and 0.2415 for screw dislocation, and 0.0673 and 0.1937 for edge dislocation at RT (*36*). Diffraction peak broadening $\Delta K$ is calculated from

$$\Delta K = -\frac{\Delta d}{d^2}, \qquad (3)$$

where $d$ (nm) is the interplanar spacing of the diffraction planes and $\Delta d$ is the full width at half maximum (FWHM) obtained from single-peak fitting of neutron diffraction data after removing the instrument peak broadening. It is evident from Equation (1) that $\Delta K^2$ follows a linear relationship with $K^2 C_{hkl}$ if $C_{hkl}$ can capture the bulk average dislocation character in the specimen measured by neutron scattering.

**Fig. 3A-D** give the modified Williamson-Hall plots of V0, V8, V16, and V25 deformed at 1173 K, respectively. $\Delta K^2$ for the samples was linear fitted with $K^2 C_{hkl}$ using the $C_{hkl}$ for both screw and edge dislocations of the $\vec{b} = \frac{a}{2} <111>$ type in a BCC crystal. A transition of dislocation character is directly captured, where in NbTaTi, a perfect fit for screw dislocations and a poor fit for edge dislocations were obtained, whereas in NbTaTiV, an opposite trend of a perfect fit for edge dislocations and a poor fit for screw dislocations were observed. Adding the analyses for the two intermediate compositions, V8 and V16, **Fig. 3E** visualizes this gradual transition from screw dislocation glide- to edge dislocation glide-dominated strength via the goodness of fitting in the modified Williamson-Hall plot using the dislocation contrast factors for screw and edge dislocations, respectively. It is shown

that the goodness of fitting for edge dislocations continuously increases with a higher concentration of V whereas the goodness of fitting for screw dislocations continuously decreases. Admittedly, while this analysis is qualitatively robust to elucidate the transition from screw to edge dislocation glide-control of strength, the detailed values of the goodness of fitting cannot be related to the fraction of screw dislocations and edge dislocations. Since the crystal consists of various other defects and lattice distortion that can also cause peak widening, using a linear mixture of $C_{hkl}$ for screw and edge dislocations with their fractions as a separate fitting variable can cause large errors and overinterpretation of data, which is therefore not conducted here.

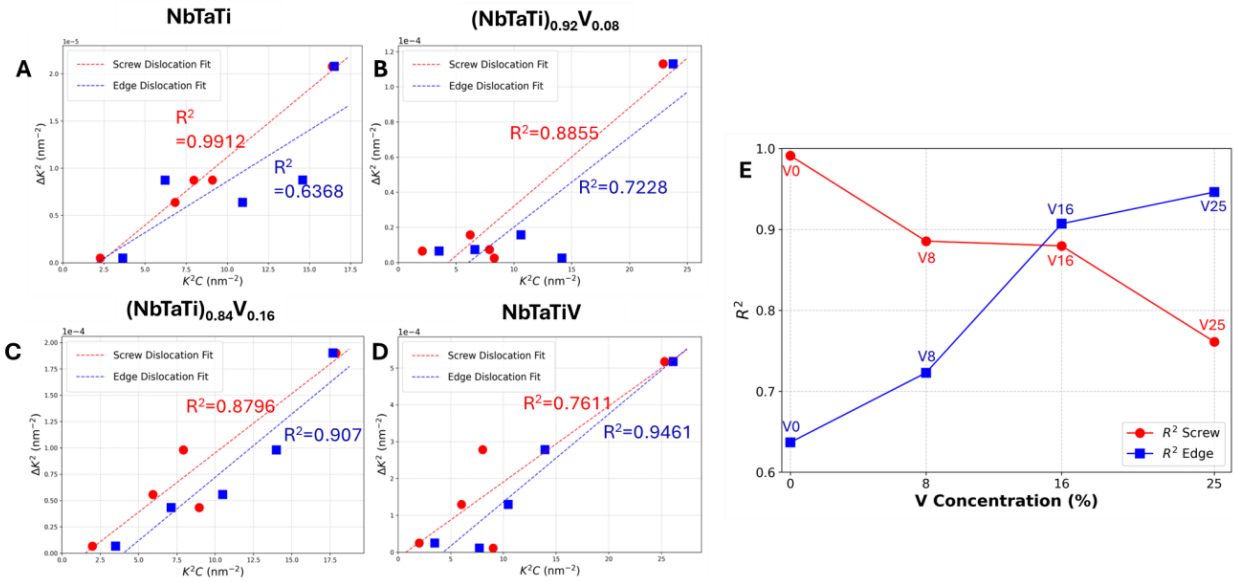

**Fig. 3 Dislocation character analysis through in-situ neutron scattering measurements.** (A-D) Modified Williamson-Hall plot for V0, V8, V16, and V25 using dislocation contrast factors for screw <1,1,1> type dislocations and edge <1,1,1> {1,1,0} type dislocations. Linear fit of $\mathbf{\Delta K^2}$ as a function of $\mathbf{K^2 C}$ was performed on all alloys for both screw and edge dislocations and the goodness of fitting, $\mathbf{R^2}$, is given in (E).

To complement the neutron analysis, detailed dislocation characterization was carried out using the diffraction contrast imaging technique via scanning transmission electron microscopy (DCI-STEM) on NbTaTi and NbTaTiV deformed at 1173 K. Since neutron scattering provides a more representative assessment of the overall dislocation character than STEM analysis, only the two extreme compositions, V0 and V25, were examined to illustrate the contrast in deformation mechanisms. Low-angle annular dark-field (LAADF) STEM imaging provides critical insights into the dislocations developed in NbTaTi and NbTaTiV RCCAs deformed under tension at 900 °C and interrupted at below 2% strain, as shown in **Fig. 4**. NbTaTi exhibits the characteristic features of classic screw dislocation-controlled plasticity. The dislocations are predominantly long, straight segments along {110} planes with frequently observed debris, which originates from the process of cross-kink unpinning that results in vacancy or interstitial dislocation loops. **Fig. 4A** shows the dislocation substructure in a high dislocation density area, where screw dislocation segments along three {110} planes were observed. At an area with a lower dislocation density depicted by **Fig. 4B**, the signature "coffee bean" contrast for Frank loops was observed as strong evidence for debris that accompanies screw dislocations. A detailed explanation and quantitative analysis of this process can be found elsewhere (*38*, *39*) while a brief summary is provided here for the reader's convenience. In RCCAs, the rugged potential energy landscape can cause frequent double kink nucleation on separate {110} planes. In addition, double kinks already formed on a screw dislocation can cross-slip to other {110} planes at intermediate temperatures. Upon the lateral migration of these cross-kinks on different {110} planes, locks are formed when they impinge on each other. Under the applied stress, the two pairs of double kinks bow out, where the simultaneous diffusion of vacancies or interstitials to the lock can facilitate a local non-conservative motion of the pinned segment to form a vacancy/interstitial loop. At the meantime, the long screw dislocation segment detaches from the lock and continues gliding. Therefore, cross-kink unpinning is a stress-assisted, thermally activated process that does not lead to a strength plateau, which is manifested by the weak plateau behavior observed in the temperature dependent yield strength of NbTaTi. This behavior is consistent with the strong screw dislocation glide control of yield strength commonly observed in non-V containing RCCAs, such as HfNbTaTiZr, $Nb_{45}Ta_{25}Hf_{15}Ti_{15}$, and NbTiZr, as discussed in the **Introduction** section.

In strong contrast is the dislocation substructure in NbTaTiV (**Fig. 4C-D**), where highly curvy dislocations with mixed characters were observed. In addition, the formation of debris is largely reduced, suggesting a fundamentally different deformation mechanism compared to that in NbTaTi. With the addition of V that introduces a large atomic size misfit and hydrostatic pressure field, the mobility of edge dislocations can sharply decrease and the mobility for screw dislocations can remain largely unaffected. The consequence is that screw dislocations and edge dislocations face similar glide barriers. This scenario is particularly like FCC CCAs with similar edge and screw dislocation glide barriers and distinct from conventional BCC alloys where the glide barrier for screw dislocations is always higher than edge dislocations. As a result, the dislocation substructure shown in **Fig. 4C-D** largely resembles that in FCC CCAs such as CrMnFeCoNi and CrCoNi deformed at elevated temperatures as shown in Refs. (*40–42*) where curvy dislocations uniformly distributed on the glide planes were observed. Specifically, the curvy morphology of dislocations manifests the local relaxation of dislocation segments according to the rugged potential energy landscape from the concentrated solid solution. Of course, partial dislocations separated by stacking faults that are frequently observed in FCC CCAs are missing in this BCC RCCA due to the high stacking fault energy in BCC systems (*43*). As a result, the deformation in NbTaTiV is controlled by a competition between edge and screw glide. The curved and tangled dislocations indicate a transition toward control of the edge character that can be increasingly strong with increased lattice distortion arising from the addition of V atoms.

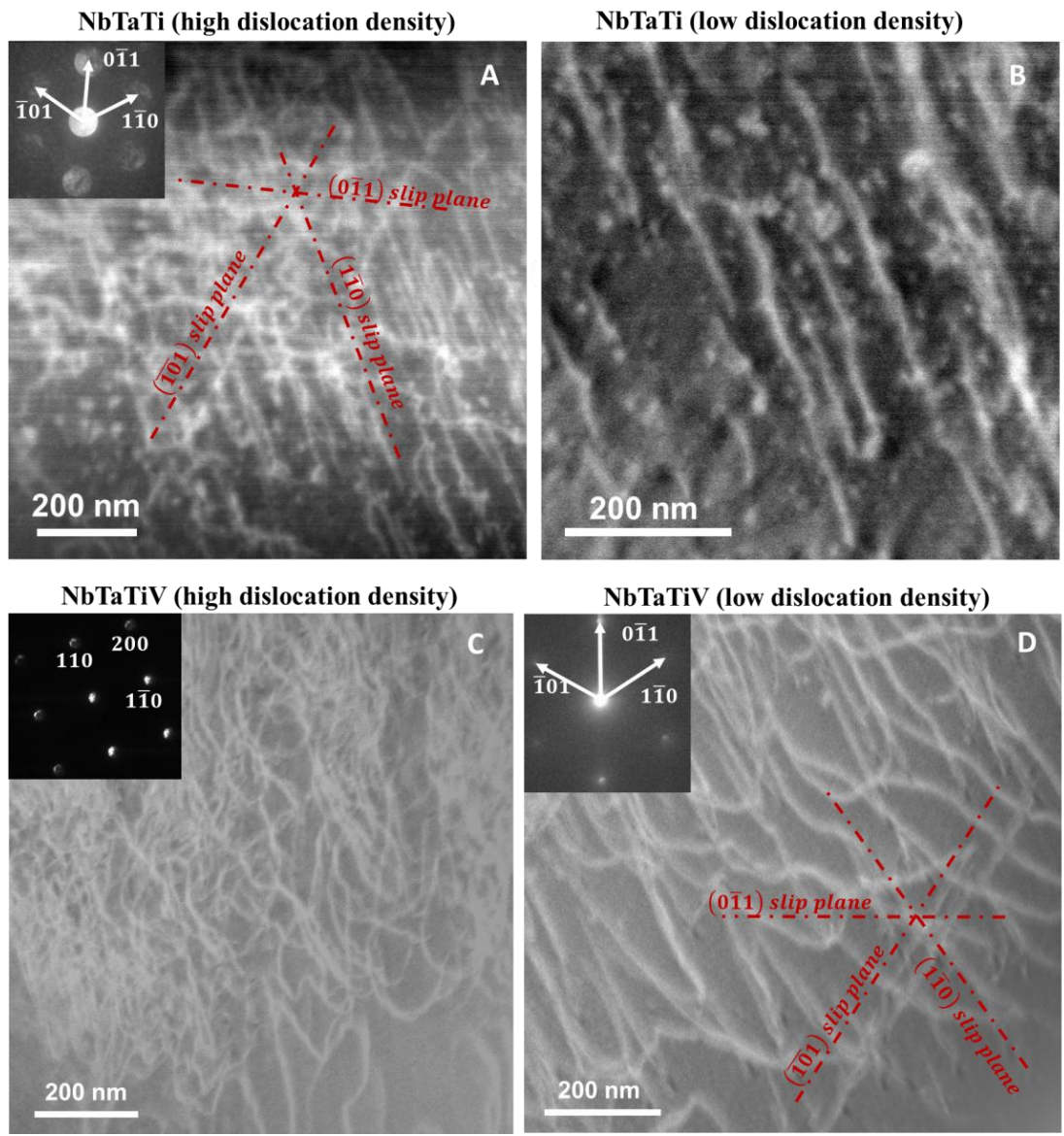

**Fig. 4 STEM-LAADF dislocation images.** (A) A high density of screw dislocations observed along the [111] zone axis in the NbTaTiV alloy deformed under tension at 1173 K and interrupted at below 2% strain. (B) A region in NbTaTi with a low dislocation density. The inset of (A) shows the corresponding diffraction pattern at a convergence semi-angle of ~1 mrad. (C) A high density of mixed dislocations observed along the [100] and [111] zone axes in the NbTaTiV alloy deformed under tension at 1173 K and interrupted at below 2% strain. (D) A region with a low dislocation density. The insets of (C-D) show the corresponding diffraction patterns obtained from selected area diffraction under conventional TEM mode before switching to DCI-STEM imaging.

To provide further support for the above arguments, MD simulations were performed to compute the flow stress of edge and dislocation glide in NbTaTi and NbTaTiV. **Fig. 5** shows the results of MD computed flow stresses for edge and screw dislocation glide, overlaid with the experimental yield strength data. It is observed that in NbTaTi, screw dislocations always possess higher barriers compared to their edge counterparts, therefore controlling the strength. On the other hand, a cross-over in the flow stress exists for NbTaTiV, where screw dislocations face stronger barriers up to ~800 K, above which edge dislocations take over to control the strength. This cross-over temperature aligns well with the location of the yield strength plateau in NbTaTiV, indicating that the onset of the plateau is correlated with the transition from screw to edge dislocation glide control. However, it is reiterated here that the thermally activated edge dislocation glide alone cannot fully rationalize the plateau and the YSA behavior, and additional strengthening mechanisms are required. Coincidentally, the magnitude of MD flow stress also coincides with the experimental yield stress, but it is highlighted that this phenomenon may be fortuitous due to the combination of two factors: the high MD strain rate overestimates the strength; and the negligence of other strengthening mechanisms such as grain boundary strengthening and dislocation forest strengthening underestimate strength.

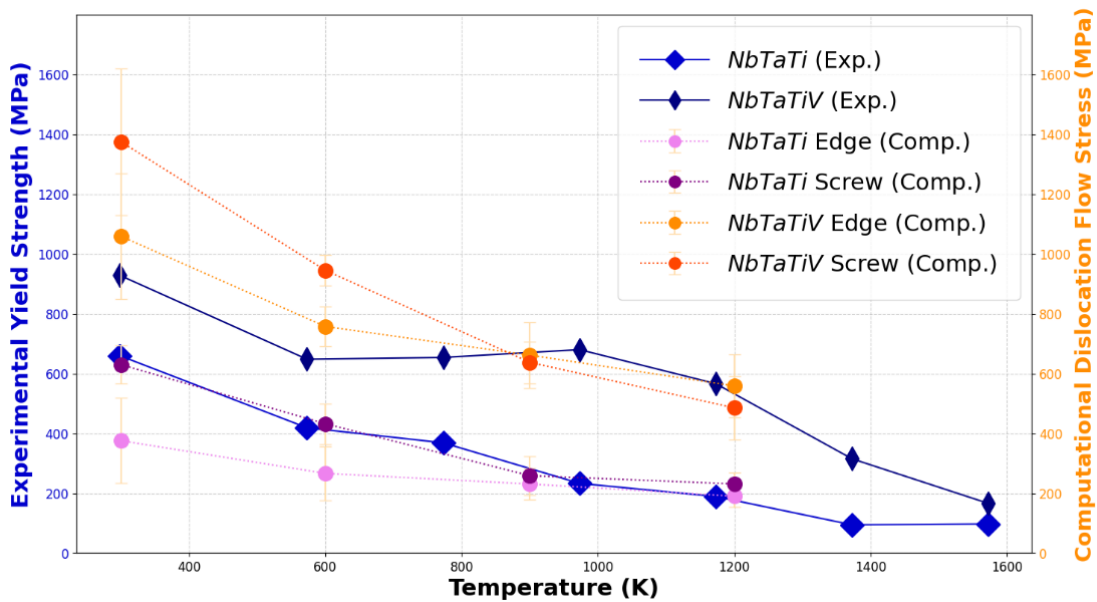

**Fig. 5 MD computed flow stresses for edge and screw dislocation glide, overlaid with the experimental yield strength data of NbTaTi and NbTaTiV alloys.** The flow stress of screw and edge strength features a cross-over at ~800 K for NbTaTiV.

## Discussion

### Phase stability of the (NbTaTi)$_{1-x}$V$_x$ RCCAs

It is important to discuss the phase stability of the (NbTaTi)$_{1-x}$V$_x$ RCCAs, as the BCC single-phase stability of this system cannot always be assumed. The calculated phase diagram (CALPHAD) calculations for all four alloy compositions are shown in **Fig. S2** to evaluate the tendency for secondary phase or oxide/nitride formation that could potentially alter the strength of the material and cause embrittlement. CALPHAD was performed with and without the consideration of O/N impurities, and the O/N levels used in this computation were derived from the inert gas fusion measurements as shown in **Table 1**. For the alloys without impurities, CALPHAD predictions indicate a low-temperature Ti-rich HCP phase in all four alloys. With the addition of V, a BCC miscibility gap can be driven by the separation of high melting point (Nb,Ta) and low melting (Ti, V) to form a BCC#1 and a BCC #2 phase. Ordered intermetallic phases can form at even lower temperatures, with B2 and Laves phases suggested by the computation. In reality, the samples are most often homogenized or annealed at a high temperature where a single-phase BCC is considered stable, followed by cooling under various rates depending on the annealing methods and furnace setup (e.g., vacuum quartz encapsulation followed by quenching, annealing under Ar protection followed by furnace cooling), Whether these phases will actually form depends on the interplay of thermodynamics, kinetics, and the thermomechanical processing history, which can only be examined by experiments. In literature, only the Ti-rich phase was reported in the NbTaTiV system (*44, 45*), whereas the presence of the BCC miscibility gap and ordered intermetallics has not yet been seen possibly due to sluggish kinetics.

It is noticeable that with the addition of V, the stability of the HCP phase is progressively suppressed to lower temperatures. In the equiatomic NbTaTiV alloy, the HCP phase formation is restricted to below ~500 K (below 0.2 T$_m$, **Fig. S2D**). This trend indicates that V additions or the according decrease of the Ti content enhance the thermodynamic stability of the BCC solid-solution phase and inhibit the transformation to the HCP phase. However, it is important to note that the O/N impurities can significantly stabilize the Ti-rich phase due to a high affinity of Group IV metals with O and N (*46, 47*). **Fig. S2E-H** show that the HCP phase can be stable up to the solidus temperature of the alloy, which encompasses all homogenization and test temperatures. Furthermore, a Ti$_2$N nitride phase can be stable at lower temperatures. Throughout the investigation, the samples were carefully handled and tested in high vacuum (except for the in-situ neutron tensile tests), and the O/N impurities were likely to originate from the stock material. These minute amounts correlate to extremely low phase fractions at elevated temperatures, although the CALPHAD computation can only address bulk samples under thermodynamic equilibrium. Possible grain boundary segregation of these phases can occur in this system despite a low overall phase fraction and has been observed in previous work (*45*). Therefore, in this investigation, faster "air cooling" was applied after annealing while the samples continued to stay in high vacuum. This is achieved by a sliding tube furnace design where the vacuum tube was immediately removed from the heat source upon the completion of annealing, which offers a cooling rate approximately ten times faster than furnace cooling to kinetically suppress the formation of the low-temperature phases. The incorporation of such design if critical to allow a pristine comparison of the mechanical properties of the RCCAs in single-phase BCC solid solutions.

**Plateau behavior in temperature-dependent yield strength**

The differences in the plateau behavior as a function of V content shown in **Fig. 2** suggests a possible transition in the rate-controlling deformation mechanism from screw dislocation motion in NbTaTi to a mixed regime, where the yield strength can be controlled by the competition between screw and edge dislocation glide. This transition can lead to a more gradual decline in strength as a function of temperature, since the edge dislocation-concentrated solid solution interactions are envisaged to be much less temperature-sensitive than double kink and cross-kink mechanisms for screw dislocations (*19*), (*20*), (*30*), (*36, 38, 39, 44–51*). It is important to determine whether this transition arises solely from the atomic size misfit effect, in which the hydrostatic pressure field introduced by V atoms, with their smaller atomic radius, interacts with edge dislocations but not with screw dislocations, as the latter generate only shear stress fields. To test this hypothesis, temperature-dependent yield strengths for the NbTaTi$_{1-x}$V$_x$ series and binary V–Nb alloys [cite] at room temperature, 573 K (onset of plateau), and 1173 K (end of plateau) are compared in **Fig. S3**. Given the similar metallic radii of Nb, Ta, and Ti (~146 pm) and their notable difference from V (134 pm) (*52*), the binary alloys effectively isolate and represent the atomic size misfit contribution. Lattice distortion clearly contributes to the overall strength trend, yet the RCCAs, especially V25, are significantly stronger than the binary alloys (*53*), indicating additional strengthening mechanisms. In addition, the YSA behavior observed in V8, V16, and V25, where the yield strength increases with increasing temperature near 773 K, cannot be explained by edge dislocation glide control alone. Since the transition from screw- to edge-dominated dislocation glide can only account for a decrease in the steepness of the strength-temperature slopes, a secondary strengthening mechanism, especially centered around edge dislocations, must be present to fully rationalize the plateau and YSA behavior. YSA is typically observed in Ni-based superalloys and B2-ordered FeAl due to antiphase boundary shearing and dynamic strain aging(*51, 54, 55*), which does not apply to the current RCCAs as no secondary phases are present and the interstitial levels for all four compositions are compatible with each other. Alternative possible explanations include: 1. The addition of V can drive the formation of chemical short-range order (CSRO) that escapes the current characterization, which requires a certain stress to destroy during the onset of plastic deformation. This CSRO can be peaked at 25% V addition and 773 K; 2. The occurrence of cross-core diffusion, where larger solute atoms diffuse to the tensile side of edge dislocations and smaller solute atoms diffuse to the compressive side of edge dislocations across the edge dislocation cores, thereby creating strengthening as the dislocations have to break through this energetically favorable environment during plastic deformation (*56–59*); and 3. The nucleation of vacancies at edge dislocation cores that forms dislocation dipoles or superjogs, creating a dragging force on the dislocations (*60*). The investigation of these additional mechanisms requires a comprehensive characterization of possible ordering as well as meticulous measurements of the equilibrium vacancy concentrations, strain rate sensitivity, and activation volumes, which are beyond the scope of the current study and referred to future work.

**Implications on RCCA design principles**

A final discussion is presented here on the implications of the observed transition in deformation mechanisms at elevated temperatures in the NbTaTiV system for future alloy design of strong and ductile RCCAs. The shift to edge dislocation glide–controlled strength, induced by the large atomic size misfit introduced through V alloying, is crucial for retaining strength at intermediate temperatures, as evidenced by the higher yield strength of NbTaTiV compared to NbTaTi. For an annealed, coarse-grained, single-phase material, NbTaTiV demonstrates excellent strength, with a yield strength comparable to that of tempered martensitic steels at room temperature and Ni-based superalloys at intermediate temperatures (*61*, *62*). Above 1373 K, NbTaTiV exhibits a yield strength exceeding that of Ni-based superalloys (*61*). Compared to commercial refractory alloys, NbTaTiV exhibits a higher yield strength than both annealed and additively manufactured C103, a Nb-based alloy containing minor additions of Hf and Zr (*63*, *64*). In addition, NbTaTiV shows a strength comparable to that of TZM, a Mo-based alloy strengthened by small additions of Ti and Zr, but is much more ductile (*65*). On the other hand, previous studies have reported excellent tensile ductility for NbTaTiV and NbTaTi. For the NbTaTiV alloy, the brittle-to-ductile transition ratio (Pugh ratio, or the ratio between the bulk modulus and the shear modulus, K/G) is 3.41, well above the critical threshold of 1.75, and it showed more than 40% tensile ductility at room temperature from the as-cast state (*22*). To date, NbTaTiV is one of the very few RCCAs that can demonstrate a combination of high-temperature strength and room-temperature tensile ductility, a critical design goal for the proposed application areas for RCCAs. Future research on NbTaTiV should leverage its remarkable tensile ductility to develop deformation-processing routes and multiphase strengthening strategies that enhance the strength–ductility synergy for high-temperature applications. Moreover, its creep properties remain to be systematically explored. Considerable potential exists to further strengthen NbTaTiV through dispersion or precipitation hardening and advanced processing methods such as additive manufacturing. Notably, recent studies have demonstrated that NbTaTiV can be additively manufactured crack-free (*66*).

In summary, the comprehensive experimental and computational work outlined in this study is believed to provide definitive evidence to reveal the mechanistic transition from screw dislocation glide- to edge dislocation glide-controlled yield strength at elevated temperatures caused by V-alloying in the NbTaTi system. The shift in deformation mechanism not only drastically increases the high-temperature strength of the alloy but also leads to a distinct intermediate-temperature yield strength plateau for better strength retention. This work highlights that targeted alloying to promote edge dislocation control represents an effective strategy for achieving superior high-temperature mechanical properties in RCCAs.

## Materials and Methods

### Alloy fabrication and processing

To investigate NbTaTi RCCAs alloyed with increasing vanadium contents, $(NbTaTi)_{1-x}V_x$ alloys with x = 0.25 (NbTaTiV), x = 0 (NbTaTi), and two intermediate compositions where x = 0.16 $(NbTaTi)_{0.84}V_{0.16}$, and x = 0.08 $(NbTaTi)_{0.92}V_{0.08}$, were fabricated by vacuum arc melting of ultrahigh-purity elements using an Arcast Arc 200 arc melter at UC Davis. Prior to arc melting, the chamber was evacuated to below $5 \times 10^{-5}$ torr and subsequently backfilled with ultrahigh-purity argon. The melted alloy button was flipped and remelted at least five times to ensure a homogeneous composition.

Following arc melting, the alloys were cold rolled crack-free to ~85% thickness reduction, reaching a final thickness of 1 mm using an IRM2050 2-hi rolling mill. Dog-bone specimens for tensile testing were then machined by electro-discharge machining (EDM) from the rolled sheets. Following rolling, the NbTaTiV and $(NbTaTi)_{0.84}V_{0.16}$ tensile samples were wrapped in tantalum foil and annealed at 1473 K for 3 h, while all other compositions were annealed at 1473 K for 5 h under high vacuum, followed by air cooling to achieve full homogenization and recrystallization. Air cooling (~100 K/min) while maintaining high vacuum was achieved through a sliding tube furnace, where the tube was removed from the heat source immediately after annealing and cooled in air with a box fan. Titanium getters were used during both melting and annealing to absorb excess oxygen in the system. The oxygen and nitrogen impurity levels in the annealed samples prior to tensile testing were measured by inert gas fusion at the IMR Test Labs, Portland, OR using the LECO OHN836 system. 5g of each specimen was used for the tests.

### Calculated phase diagram (CALPHAD) computations

To understand the theoretical phase stability of the alloys, CALPHAD computation was performed on V0, V8, V16, and V25 using the CompuTherm PanRHEA database. The computation was carried out from room temperature to 3000 K. To examine the effect of O and N impurities on phase stability, separate computations were conducted using the O and N levels measured by inert gas fusion.

### Room and high temperature tensile testing

Room temperature tensile tests were conducted in air on an MTS Model 810 servohydraulic machine. Elevated-temperature tensile tests were performed between 573–1573 K, in 200 K intervals, under a high vacuum better than $5 \times 10^{-5}$ torr using an Instron 1331 hydraulic testing machine equipped with a custom-built vacuum furnace. This Mo-element furnace is capable of high-temperature tensile testing up to 1873 K and is equipped with a high-temperature thermocouple positioned near the gauge section of the tensile specimen to precisely measure the actual temperature. The tensile specimens were heated at 15 K/min to the target temperature, stabilized at the temperature for 5 min and then deformed at a strain rate of $5 \times 10^{-4}$ s$^{-1}$. Since only the yield strengths of these RCCAs are of interest, all tensile tests were interrupted at below 2% tensile strain for further electron microscopy imaging. The samples were cooled down under load to preserve the dislocation substructure. Water-cooled TZM grips and furnace elements ensured fast cooling that brought down the temperature from the tensile tests rapidly to minimize annealing of the microstructure.

### Scanning transmission electron microscopy (STEM) imaging

For STEM imaging, samples were mechanically polished to a thickness of 80-100 μm using 1200 grit carbide sandpaper and 3 mm discs were cut with a TEM sample punch. The discs were then electropolished using a Fischione 110 twin jet electropolisher using an electrolyte of 10 vol% sulfuric acid in methanol. A mixture of methanol and liquid nitrogen was used as the coolant to maintain the temperature between −30 °C and −20 °C. The same electropolishing procedure was followed for preparing samples for scanning electron microscopy (SEM) and energy dispersive X-ray spectroscopy (EDS). STEM imaging was performed an FEI F20 UT Tecnai under 200 kV at the National Center for Electron Microscopy (NCEM), Lawrence Berkeley National Laboratory (LBNL). SEM and EDS analyses were performed using a ThermoFisher Quattro S Environmental SEM) at the Advanced Materials Characterization and Testing Laboratory (AMCaT), UC Davis. SEM imaging was performed on the annealed undeformed samples before the tensile testing at 20 kV with a 10 mm working distance.

### In-situ heating/tension neutron scattering experiments

Neutron diffraction measurements were performed during in-situ heating and tensile testing on the VULCAN Engineering Diffractometer, Spallation Neutron Source (SNS), Oak Ridge

National Lab (ORNL). The time-of-flight neutron diffraction covers a wide range of d-spacing without rotating the tensile specimen and it can directly extract microstrains for different {hkl} planes in one measurement. Uniaxial tensile testing was conducted at 5x $10^{-6}$ s$^{-1}$ strain rate at 1173 K on the MTS load frame that can go up to +/-100KN tension with induction heating in air. A slower tensile strain rate was used to allow ample time for neutron data acquisition. Large, annealed dog bone tensile specimens with a total length of 100 mm, gauge length of 15 mm, and thickness of 2.5-3 mm were used to ensure adequate and homogeneous heating. A K-type thermocouple was spot-welded on each sample to measure the real-time temperature of the sample gauge and control the power of furnace heating. Previously, the temperature variation along the gauge section of the sample geometry was benchmarked to by below 10 K. All alloys were tested following a four-step loading sequence. Initially, tensile specimens were elastically loaded and unloaded at room temperature to obtain the elastic constants, followed by a heating ramp under load control with a 5 min hold at target temperature to ensure thermal stability, subsequent high temperature tensile testing. Tensile tests were then terminated at ~2% strain, after which specimens were cooled to room temperature.

On VULCAN, diffraction patterns were collected simultaneously in the axial and transverse directions of the tensile specimen using two He linear position-sensitive are detectors that are located at ±90° relative to the incident neutron beam while the specimen was continuously tensioned at 45° relative to the incident beam. An incident beam slit of 8 mm x 2.5 mm and a 5 mm receiving collimator was used. The chopper frequency was set at 30 Hz that could pass through high neutron flux while providing sufficient d-spacing coverage of ~0.5-2.5 angstroms at the ±90° detector banks. The continuously collected neutron diffraction data and sample environment data (e.g. stress, strain, temperature) were chopped into one- to two-minute bins using VDRIVE (*67*) so each bin of data represented the averaged information within that time interval. This is in contrast to the conventional step-loading or interrupted-loading measurements that could introduce stress relaxation during strain holding for neutron data collection. The single peak fitting was performed for the first six diffraction peaks using VDRIVE (*67*) and peak broadening data was further analyzed to characterize the lattice constants, elastic moduli, and character of dislocations for the samples. Neutron data from yield to ~1% plastic strain was used to analyze the dislocations that control the yield strength of RCCAs. Other dislocation substructures, such as twinning and kink band formation that occur at a much later stage of deformation, are therefore not included in the neutron analysis.

**Molecular dynamics (MD) simulations**
MD simulations were performed with the Large-scale Atomic/Molecular Massively Parallel Simulator (LAMMPS) (*68*). A modified Embedded Atom Method (MEAM) potential was used for NbTaTi and NbTaTiV systems (*69*). We simulated single dislocations using a periodic-array-of-dislocations (PAD) setup following prior work (*34, 70*). Edge and screw a/2⟨111⟩ dislocations gliding on (110) were modeled in BCC NbTaTi and NbTaTiV where atomic species were randomly distributed. The simulation box axes are $x[11\bar{2}]$, $y[111]$, and $z[1\bar{1}0]$ for an edge dislocation and $x[111]$, $y[11\bar{2}]$, and $z[1\bar{1}0]$ for a screw dislocation. The simulation box contains about $2\times10^5$ atoms with size ~80×250×160 Å. Periodic boundary conditions (PBCs) are applied along the *x* direction, free boundaries along the *z* direction, and either standard PBCs (edge) or shifted PBCs (screw) along the *y* direction (*71*). The top and bottom five atomic layers were fixed as boundary regions. A small displacement was applied to the top region to produce a nominal strain rate of $10^7 s^{-1}$. Flow stresses were extracted from the peak values of the stress-strain curve. All simulations used a 1 fs timestep with a Nosé–Hoover thermostat (*72*) over 300–1200 K.

**References**

1. A. Kulkarni, A. James, A. Kamel, Advanced Materials and Manufacturing Technology Developments for Extreme Environment Gas Turbine Applications. *Matls. Perf. Charact.* **10**, 146–160 (2021).

2. O. N. Senkov, D. B. Miracle, K. J. Chaput, J.-P. Couzinie, Development and exploration of refractory high entropy alloys—A review. *Journal of Materials Research* **33**, 3092–3128 (2018).

3. J. H. Perepezko, The Hotter the Engine, the Better. *Science* **326**, 1068–1069 (2009).

4. M. C. Gao, C. S. Carney, Ö. N. Doğan, P. D. Jablonksi, J. A. Hawk, D. E. Alman, Design of Refractory High-Entropy Alloys. *JOM* **67**, 2653–2669 (2015).

5. O. N. Senkov, G. B. Wilks, D. B. Miracle, C. P. Chuang, P. K. Liaw, Refractory high-entropy alloys. *Intermetallics* **18**, 1758–1765 (2010).

6. A. Gupta, R. Raj, G. Shankar, S. Suwas, S.-H. Choi, Microstructural stability and mechanical properties of the non-equiatomic NbCrTiMoVHf refractory complex concentrated alloy. *J Mater Sci* **60**, 11693–11717 (2025).

7. C. Liang, X. Deng, C. Wang, Y. Xie, X. He, X. Wu, Y. Deng, Achieving prominent strength-ductility trade-off and ultrahigh specific strength in novel Ti-V-Al-Zr-Nb LRCCA via inducing HCP precipitates. *Materials Science and Engineering: A* **947**, 149205 (2025).

8. O. N. Senkov, S. Gorsse, D. B. Miracle, High temperature strength of refractory complex concentrated alloys. *Acta Materialia* **175**, 394–405 (2019).

9. D. B. Miracle, O. N. Senkov, C. Frey, S. Rao, T. M. Pollock, Strength *vs* temperature for refractory complex concentrated alloys (RCCAs): A critical comparison with refractory BCC elements and dilute alloys. *Acta Materialia* **266**, 119692 (2024).

10. F. G. Coury, M. Kaufman, A. J. Clarke, Solid-solution strengthening in refractory high entropy alloys. *Acta Materialia* **175**, 66–81 (2019).

11. S. I. Rao, C. Woodward, B. Akdim, Solid solution softening and hardening in binary BCC alloys. *Acta Materialia* **243**, 118440 (2023).

12. G. Taylor, Thermally-activated deformation of BCC metals and alloys. *Progress in Materials Science* **36**, 29–61 (1992).

13. M. I. Wood, G. Taylor, Niobium—an athermal plateau in the low-temperature yield stress. *Philosophical Magazine A* **56**, 329–342 (1987).

14. M. I. Wood, "The deformation of bcc alloys," thesis, University of Oxford (1982).

15. R. J. Arsenault, "Low Temperature of Deformation of bcc Metals and Their Solid-Solution Alloys†" in *Treatise on Materials Science & Technology*, R. J. Arsenault, Ed. (Elsevier, 1975; https://www.sciencedirect.com/science/article/pii/B9780123418067500088)vol. 6 of *Plastic Deformation of Materials*, pp. 1–99.

16. M. R. Jones, P. Garg, I. J. Beyerlein, Critical temperatures for ductile to brittle transition and athermal strength in refractory metals. *Journal of Materials Research and Technology* **38**, 6229–6243 (2025).

17. L. H. Mills, M. G. Emigh, C. H. Frey, N. R. Philips, S. P. Murray, J. Shin, D. S. Gianola, T. M. Pollock, Temperature-dependent tensile behavior of the HfNbTaTiZr multi-principal element alloy. *Acta Materialia* **245**, 118618 (2023).

18. S. I. Rao, C. Woodward, B. Akdim, E. Antillon, T. A. Parthasarathy, O. N. Senkov, Estimation of diffusional effects on solution hardening at high temperatures in single phase compositionally complex body centered cubic alloys. *Scripta Materialia* **172**, 135–137 (2019).

19. F. Maresca, W. A. Curtin, Theory of screw dislocation strengthening in random BCC alloys from dilute to "High-Entropy" alloys. *Acta Materialia* **182**, 144–162 (2020).

20. S. I. Rao, C. Woodward, B. Akdim, O. N. Senkov, D. Miracle, Theory of solid solution strengthening of BCC Chemically Complex Alloys. *Acta Materialia* **209**, 116758 (2021).

21. C. Lee, F. Maresca, R. Feng, Y. Chou, T. Ungar, M. Widom, K. An, J. D. Poplawsky, Y.-C. Chou, P. K. Liaw, W. A. Curtin, Strength can be controlled by edge dislocations in refractory high-entropy alloys. *Nat Commun* **12**, 5474 (2021).

22. C. Guo, Y. Xing, P. Wu, R. Qu, K. Song, F. Liu, Super tensile ductility in an as-cast TiVNbTa refractory high-entropy alloy. *Progress in Natural Science: Materials International* **34**, 1076–1084 (2024).

23. O. N. Senkov, J. M. Scott, S. V. Senkova, F. Meisenkothen, D. B. Miracle, C. F. Woodward, Microstructure and elevated temperature properties of a refractory TaNbHfZrTi alloy. *J Mater Sci* **47**, 4062–4074 (2012).

24. O. N. Senkov, J. M. Scott, S. V. Senkova, D. B. Miracle, C. F. Woodward, Microstructure and room temperature properties of a high-entropy TaNbHfZrTi alloy. *Journal of Alloys and Compounds* **509**, 6043–6048 (2011).

25. L. Lilensten, J.-P. Couzinié, L. Perrière, A. Hocini, C. Keller, G. Dirras, I. Guillot, Study of a bcc multi-principal element alloy: Tensile and simple shear properties and underlying deformation mechanisms. *Acta Materialia* **142**, 131–141 (2018).

26. J.-Ph. Couzinié, L. Lilensten, Y. Champion, G. Dirras, L. Perrière, I. Guillot, On the room temperature deformation mechanisms of a TiZrHfNbTa refractory high-entropy alloy. *Materials Science and Engineering: A* **645**, 255–263 (2015).

27. Kink bands promote exceptional fracture resistance in a NbTaTiHf refractory medium-entropy alloy. https://doi.org/10.1126/science.adn2428.

28. G. Sahragard-Monfared, C. H. Belcher, S. Bajpai, M. Wirth, A. Devaraj, D. Apelian, E. J. Lavernia, R. O. Ritchie, A. M. Minor, J. C. Gibeling, C. Zhang, M. Zhang, Tensile creep behavior of the Nb45Ta25Ti15Hf15 refractory high entropy alloy. *Acta Materialia* **272**, 119940 (2024).

29. S. I. Rao, W. Wang, M. Asta, R. O. Ritchie, M. Zhang, Rationalization of the tensile creep behavior of the Nb45Ta25Ti15Hf15 *bcc* refractory complex concentrated alloy using Rao-Suzuki screw dislocation glide model. *Scripta Materialia* **265**, 116752 (2025).

30. R. R. Eleti, N. Stepanov, N. Yurchenko, S. Zherebtsov, F. Maresca, Cross-kink unpinning controls the medium- to high-temperature strength of body-centered cubic NbTiZr medium-entropy alloy. *Scripta Materialia* **209**, 114367 (2022).

31. B. Akdim, C. Woodward, S. Rao, E. Antillon, Predicting core structure variations and spontaneous partial kink formation for ½<111> screw dislocations in three BCC NbTiZr alloys. *Scripta Materialia* **199**, 113834 (2021).

32. S. I. Rao, B. Akdim, A simpler technique for determining substitutional solute-screw dislocation interaction energies in BCC structures useful for estimating solid solution screw strengthening. *Acta Materialia* **269**, 119820 (2024).

33. B. Yin, F. Maresca, W. A. Curtin, Vanadium is an optimal element for strengthening in both fcc and bcc high-entropy alloys. *Acta Materialia* **188**, 486–491 (2020).

34. J. Li, H. Xu, Dislocation properties in BCC refractory compositionally complex alloys from atomistic simulations. *Computational Materials Science* **253**, 113859 (2025).

35. C. Lee, G. Song, M. C. Gao, R. Feng, P. Chen, J. Brechtl, Y. Chen, K. An, W. Guo, J. D. Poplawsky, S. Li, A. T. Samaei, W. Chen, A. Hu, H. Choo, P. K. Liaw, Lattice distortion in a strong and ductile refractory high-entropy alloy. *Acta Materialia* **160**, 158–172 (2018).

36. Temperature dependence of elastic and plastic deformation behavior of a refractory high-entropy alloy | Science Advances. https://www.science.org/doi/full/10.1126/sciadv.aaz4748.

37. A. Borbély, J. Dragomir-Cernatescu, G. Ribárik, T. Ungár, Computer program ANIZC for the calculation of diffraction contrast factors of dislocations in elastically anisotropic cubic, hexagonal and trigonal crystals. *J Appl Cryst* **36**, 160–162 (2003).

38. A. Ghafarollahi, W. A. Curtin, Screw-controlled strength of BCC non-dilute and high-entropy alloys. *Acta Materialia* **226**, 117617 (2022).

39. X. Zhou, S. He, J. Marian, Cross-kinks control screw dislocation strength in equiatomic bcc refractory alloys. *Acta Materialia* **211**, 116875 (2021).

40. G. Sahragard-Monfared, M. Zhang, T. M. Smith, A. M. Minor, E. P. George, J. C. Gibeling, The influence of processing methods on creep of wrought and additively manufactured CrCoNi multi-principal element alloys. *Acta Materialia* **261**, 119403 (2023).

41. M. Zhang, E. P. George, J. C. Gibeling, Tensile creep properties of a CrMnFeCoNi high-entropy alloy. *Scripta Materialia* **194**, 113633 (2021).

42. M. Zhang, E. P. George, J. C. Gibeling, Elevated-temperature Deformation Mechanisms in a CrMnFeCoNi High-Entropy Alloy. *Acta Materialia* **218**, 117181 (2021).

43. R. J. Wasilewski, B.C.C. stacking fault energies. *Scripta Metallurgica* **1**, 45–47 (1967).

44. D. G. Kalali, K. S. Kumar, S. Anilkumar, K. Guruvidyathri, K. N. Kishore, J. Joardar, K. V. Rajulapati, Effect of Ti on the microstructure and mechanical properties of equiatomic NbTaTi medium-entropy alloy. *International Journal of Refractory Metals and Hard Materials* **133**, 107391 (2025).

45. O. N. Senkov, J. Gild, T. M. Butler, Microstructure, mechanical properties and oxidation behavior of NbTaTi and NbTaZr refractory alloys. *Journal of Alloys and Compounds* **862**, 158003 (2021).

46. H. Okamoto, N-Ti (Nitrogen-Titanium). *J. Phase Equilib. Diffus.* **34**, 151–152 (2013).

47. H. Okamoto, O-Ti (Oxygen-Titanium). *J. Phase Equilib. Diffus.* **32**, 473–474 (2011).

48. R. Labusch, A Statistical Theory of Solid Solution Hardening. *physica status solidi (b)* **41**, 659–669 (1970).

49. R. L. Fleischer, Substitutional solution hardening. *Acta Metallurgica* **11**, 203–209 (1963).

50. S.-P. Wang, E. Ma, J. Xu, New ternary equi-atomic refractory medium-entropy alloys with tensile ductility: Hafnium versus titanium into NbTa-based solution. *Intermetallics* **107**, 15–23 (2019).

51. D. Barba, A. J. Egan, Y. Gong, M. J. Mills, R. C. Reed, "Rationalisation of the Micromechanisms Behind the High-Temperature Strength Limit in Single-Crystal Nickel-Based Superalloys" in *Superalloys 2020*, S. Tin, M. Hardy, J. Clews, J. Cormier, Q. Feng, J. Marcin, C. O'Brien, A. Suzuki, Eds. (Springer International Publishing, Cham, 2020), pp. 260–272.

52. Chemistry of the Elements, *ScienceDirect*. https://www.sciencedirect.com/book/9780750633659/chemistry-of-the-elements.

53. N. Iwao, T. Kainuma, T. Suzuki, R. Watanabe, Mechanical properties of vanadium-base binary alloys. *Journal of the Less Common Metals* **83**, 205–217 (1982).

54. O. Calonne, A. Fraczkiewicz, F. Louchet, Yield strength anomaly in b2-ordered FeAl alloys: role of boron. *Scripta Materialia* **43**, 69–75 (2000).

55. C. Kumar, P. Kumar, Analyzing dynamic strain aging and yield strength anomaly in IN740H, a nickel-based superalloy comprising low volume fraction of γ′. *Materialia* **33**, 102028 (2024).

56. M. R. Niazi, W. A. Curtin, Strengthening of edge prism dislocations in Mg–Zn by cross-core diffusion. *Modelling Simul. Mater. Sci. Eng.* **32**, 065007 (2024).

57. E. N. Epperly, R. B. Sills, Comparison of continuum and cross-core theories of dynamic strain aging. *Journal of the Mechanics and Physics of Solids* **141**, 103944 (2020).

58. M. A. Soare, W. A. Curtin, Solute strengthening of both mobile and forest dislocations: The origin of dynamic strain aging in fcc metals. *Acta Materialia* **56**, 4046–4061 (2008).

59. W. A. Curtin, D. L. Olmsted, L. G. Hector, A predictive mechanism for dynamic strain ageing in aluminium–magnesium alloys. *Nature Mater* **5**, 875–880 (2006).

60. S. He, X. Zhou, D. Mordehai, J. Marian, Thermal super-jogs control the high-temperature strength plateau in Nb-Mo-Ta-W alloys. *Acta Materialia* **244**, 118539 (2023).

61. Turbine Aerodynamics, Heat Transfer, Materials, and Mechanics, *Progress in Astronautics and Aeronautics*. https://arc.aiaa.org/doi/book/10.2514/4.102660.

62. G. Krauss, Martensite in steel: strength and structure. *Materials Science and Engineering: A* **273–275**, 40–57 (1999).

63. K. Joshi, P. Kumar, Strength Behavior of Niobium-Based Refractory Systems. *JOM* **76**, 6277–6301 (2024).

64. E. Brizes, J. Milner, E. Young-Dohe, A. Garg, R. Noebe, Effect of Temperature on Tensile Properties of Laser Powder Bed Fusion Additively Manufactured Niobium Alloy C103. *Metall Mater Trans A* **56**, 928–941 (2025).

65. T. Mrotzek, U. Martin, A. Hoffmann, High temperature deformation behavior of the molybdenum alloy TZM. *J. Phys.: Conf. Ser.* **240**, 012079 (2010).

66. A. C. Araujo, A. Cantarel, F. Chabert, A. Korycki, P. Olivier, F. Schmidt, *Material Forming: ESAFORM 2024* (Materials Research Forum LLC, 2024).

67. VDRIVE manual.pdf. https://sns.gov/sites/default/files/VDRIVE%20manual.pdf.

68. A. P. Thompson, H. M. Aktulga, R. Berger, D. S. Bolintineanu, W. M. Brown, P. S. Crozier, P. J. in 't Veld, A. Kohlmeyer, S. G. Moore, T. D. Nguyen, R. Shan, M. J. Stevens, J. Tranchida, C. Trott, S. J. Plimpton, LAMMPS - a flexible simulation tool for particle-based materials modeling at the atomic, meso, and continuum scales. *Computer Physics Communications* **271**, 108171 (2022).

69. M. S. Nitol, M. J. Echeverria, K. Dang, M. I. Baskes, S. J. Fensin, New modified embedded-atom method interatomic potential to understand deformation behavior in VNbTaTiZr refractory high entropy alloy. *Computational Materials Science* **237**, 112886 (2024).

70. "Chapter 88 Dislocation–Obstacle Interactions at the Atomic Level" in *Dislocations in Solids* (Elsevier, 2009; https://www.sciencedirect.com/science/chapter/bookseries/abs/pii/S1572485909015010)vol. 15, pp. 1–90.

71. D. Rodney, Molecular dynamics simulation of screw dislocations interacting with interstitial frank loops in a model FCC crystal. *Acta Materialia* **52**, 607–614 (2004).

72. D. Evans, B. Holian, The Nose–Hoover thermostat. *The Journal of Chemical Physics* **83**, 4069–4074 (1985).

**Acknowledgments**

**Funding:** The work was supported by the grant DE-SC0025388 funded by the U.S. Department of Energy, Office of Science. A portion of this study was carried out at the UC Davis Advanced Material Characterization and Testing (AMCaT) Facility. Funding for the Thermo Fisher Quattro S was provided by the National Science Foundation Grant No. MRI-1725618. Work at the National Center for Electron Microscopy, Molecular Foundry was supported by the Office of Science, Office of Basic Energy Sciences, of the U.S. Department of Energy under Contract No. DE-AC02-05CH11231. A portion of this research used resources at the Spallation Neutron Source, a DOE Office of Science User Facility operated by the Oak Ridge National Laboratory. The beam time was allocated to VULCAN Engineering Materials Diffractometer on proposal numbers IPTS-34010 and IPTS-35022.

**Author contributions:**
Conceptualization: MZ
Methodology: TZ, DY, JL, YC, HX, MZ
Investigation: TZ, AN, DY, JP, JL, MK, LA, YC, MZ
Visualization: TZ, DY, JL, MZ
Formal Analysis: TZ, DY, JL, MZ
Data Curation: TZ, DY, JL, MZ
Supervision: PKL, HX, MZ
Writing—original draft: TZ, MZ
Writing—review & editing: DY, PKL, HX, MZ
Project Administration: DY, PKL, HX, MZ
Funding Acquisition: MZ
Resources: DY, PKL, HX, MZ

**Competing interests:** Authors declare that they have no competing interests.

**Data and materials availability:** All data are available in the main text or the supplementary materials.

# Supplementary Materials for

# Mechanistic Transition from Screw to Edge Dislocation Glide Enhances High-Temperature Strength in Refractory Complex Concentrated Alloys

Tamanna Zakia *et al.*

*Corresponding author. Mingwei Zhang,  mwwzhang@ucdavis.edu

**This PDF file includes:**

Supplementary Text
Figs. S1 to S3

**Supplementary Text**

Calculation of intrinsic lattice distortion

The intrinsic lattice distortion can be expressed as:

$$\overline{u^D} = \sqrt{\sum_i^n f_i (d_i^{\text{eff}} - \bar{d})^2}, \qquad (1)$$

where $\bar{d}$ is the average first nearest-neighbor (1NN) interatomic distance of the alloy (in this case, $\frac{\sqrt{3}a}{2}$, where $a$ is the lattice constant), and $d_i^{\text{eff}}$ is the effective 1NN distance for the $i^{\text{th}}$ element in the alloy environment. The atomic fraction ($f_j$) is introduced as a weighting average to account for the compositional asymmetry among the constituent elements. For the continuous line plot in **Fig. S1**, $\bar{d}$ is estimated on a rule-of-mixture basis. The intrinsic lattice distortion ($\bar{u}_i^D$) represents the atomic displacement from ideal lattice sites and reflects the degree of local strain within the solid-solution matrix. The effective interatomic distance is determined by:

$$d_i^{\text{eff}} = \sum_j^n f_j \left(1 + \frac{\Delta V_{ij}}{V_i}\right)^{1/3} d_i, \qquad (2)$$

where $V_i$ and $d_i$ denotes the atomic volume and the interatomic distance for the pure $i^{\text{th}}$ element, respectively. $\Delta V_{ij}$ represents the change in atomic volume when element $i$ interacts with element $j$. The theoretical model considers both the atomic size and volume mismatch among constituent elements, where the effective lattice constant of each element is modified by its neighboring atomic environment. The $\Delta V_{ij}$ values used in this work were obtained from Lee *et al.*(*37*), where $\Delta V_{ij}$ for all six binary combinations within the Nb–Ta–Ti–V system were computed by density functional theory (DFT) simulations.

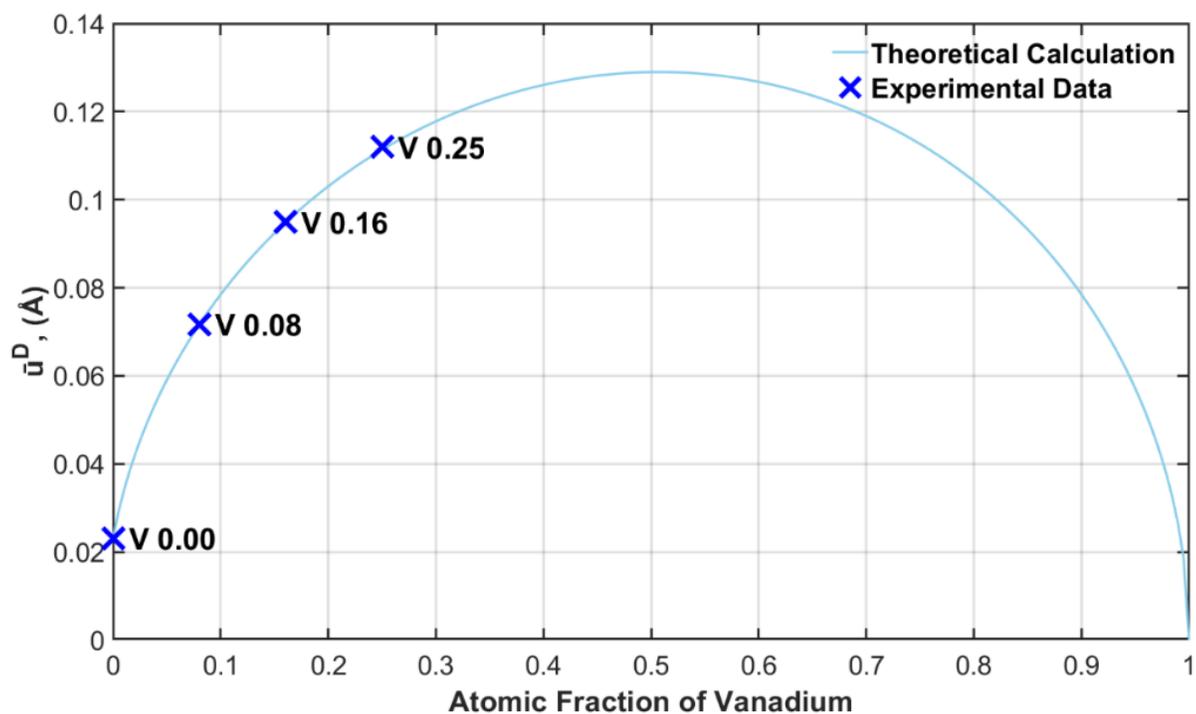

**Fig. S1. Theoretical lattice distortion,** calculated from DFT-derived interaction volumes, as a function of V content in the pseudo-binary $(NbTaTi)_{1-x}V_x$ system. Experimental lattice distortion values were calculated using the same formula but with the lattice constants derived from neutron diffraction measurements for compositions V0, V8, V16, and V25.

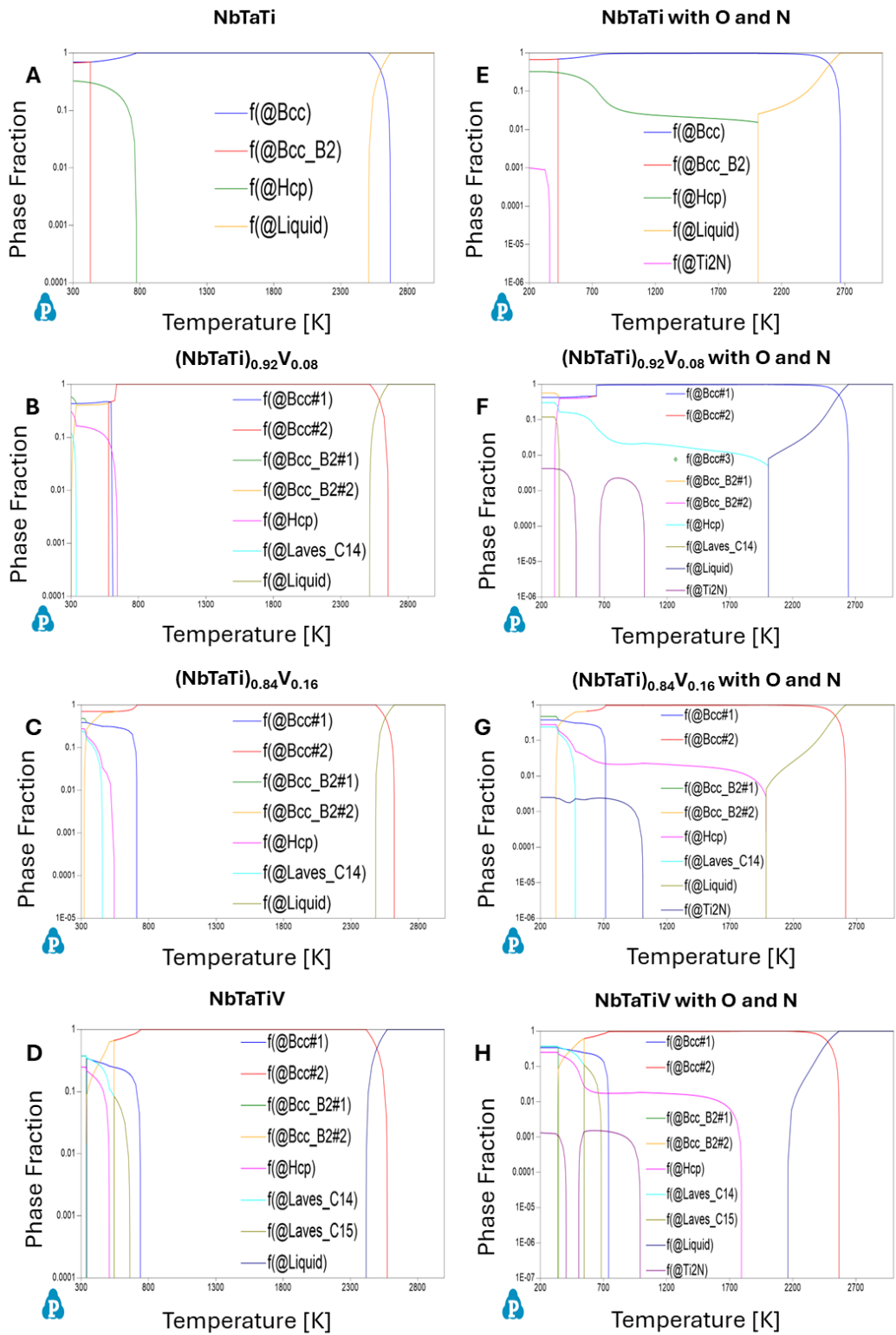

**Fig. S2 CALPHAD results for all four compositions with and without O/N impurities.** The atomic concentrations of O and N used in the computation are from inert gas fusion measurements.

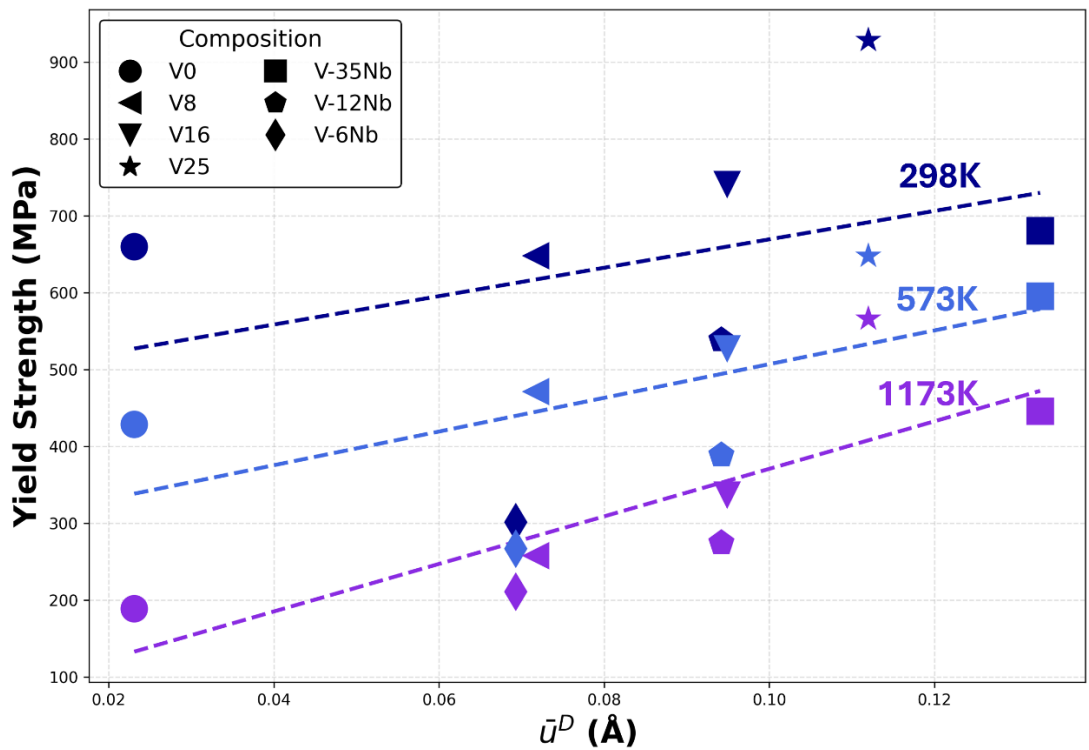

**Fig. S3 Comparison of temperature-dependent yield strengths of NbTaTi$_{1-x}$V$_x$ and binary V-Nb alloys (V$_{65}$Nb$_{35}$, V$_{88}$Nb$_{12}$, and V$_{94}$Nb$_6$, all in atomic percent) as a function of lattice distortion dominated by atomic size misfit.** The equiatomic NbTaTiV alloy exhibits a strength exceeding the overall trend, indicating evidence for the presence of additional strengthening mechanisms. The data for the binary V-Nb alloys were taken from Ref. (53).